\documentclass[runningheads]{llncs}

\usepackage{graphicx}
\usepackage{xcolor}
\usepackage{booktabs}
\usepackage{amssymb,amsmath}
\usepackage{pifont}
\usepackage{multirow}
\usepackage{appendix}

\def\eg{\textit{e.g.}~}
\def\ie{\textit{i.e.}~}

\newcommand{\V}[1]{{\boldsymbol{#1}}}

\begin{document}
\title{U-Net Transformer: Self and Cross Attention for Medical Image Segmentation}
\titlerunning{U-Transformer: Self and Cross Attention for Medical Image Segmentation}
%
\author{Olivier Petit\inst{1,2} \and
Nicolas Thome\inst{1} \and
Clement Rambour\inst{1} \and
Luc Soler\inst{2}}
\authorrunning{O. Petit et al.}
%
\institute{CEDRIC - Conservatoire National des Arts et Metiers, Paris, France \\
\and Visible Patient SAS, Strasbourg, France \\
\email{olivier.petit@visiblepatient.com}}
\maketitle              
\begin{abstract}
Medical image segmentation remains particularly challenging for complex and low-contrast anatomical structures. In this paper, we introduce the U-Transformer network, which combines a U-shaped architecture for image segmentation with self- and cross-attention from Transformers. U-Transformer overcomes the inability of U-Nets to model long-range contextual interactions and spatial dependencies, which are arguably crucial for accurate segmentation in challenging contexts. To this end, attention mechanisms are incorporated at two main levels: a self-attention module leverages global interactions between encoder features, while cross-attention in the skip connections allows a fine spatial recovery in the U-Net decoder by filtering out non-semantic features. Experiments on two abdominal CT-image datasets show the large performance gain brought out by U-Transformer compared to U-Net and local Attention U-Nets. We also highlight the importance of using both self- and cross-attention, and the nice interpretability features brought out by U-Transformer.

\end{abstract}
\section{Introduction}
\label{section:intro}

Organ segmentation is of crucial importance in medical imaging and computed-aided diagnosis, \eg for radiologists to assess physical changes in response to a treatment or for computer-assisted interventions.

Currently, state-of-the-art methods rely on Fully Convolutional Networks (FCNs), such as U-Net and variants~\cite{unet,3d-unet,v-net,unet++}. U-Nets use an encoder-decoder architecture: the encoder extracts high-level semantic representations by using a cascade of convolutional layers, while the decoder leverages skip connections to re-use high-resolution feature maps from the encoder in order to recover lost spatial information from high-level representations.

\begin{figure}[t!]
    \centering
    \includegraphics[width=\textwidth]{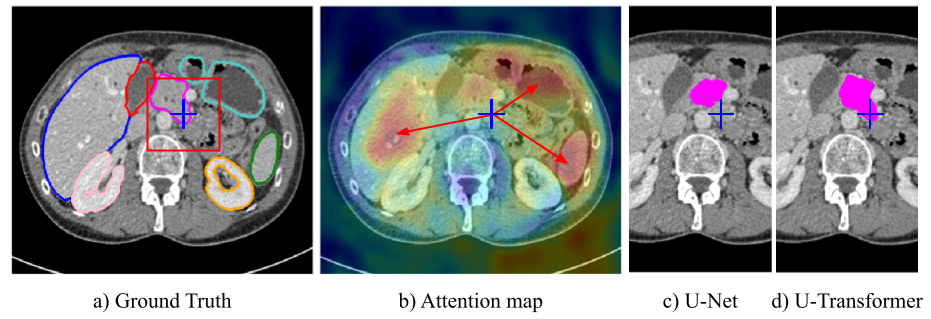}
    \caption{Global context is crucial for complex organ segmentation but cannot be captured by vanilla U-Nets with a limited receptive field, \ie blue cross region in a) with failed segmentation in c). The proposed U-Transformer network represents full image context by means of attention maps b), which leverage long-range interactions with other anatomical structures to properly segment the complex pancreas region in d).}
    \label{fig:unet_rf}
    \vspace{-0.6cm}
\end{figure}

Despite their outstanding performances, FCNs suffer from conceptual limitations in complex segmentation tasks, \eg when dealing with local visual ambiguities and low contrast between organs. This is illustrated in Fig~\ref{fig:unet_rf}a) for segmenting the blue cross region corresponding to the pancreas with U-Net: the limited receptive field framed in red does not capture sufficient contextual information, making the segmentation fail, see Fig~\ref{fig:unet_rf}c).

In this paper, we introduce the U-Transformer network, which leverages the strong abilities of transformers~\cite{Vaswani_NIPS_2017} to model long-range interactions and spatial relationships between anatomical structures. U-Transformer keeps the inductive bias of convolution by using a U-shaped architecture, but introduces attention mechanisms at two main levels, which help to interpret the model decision. Firstly, a self-attention module leverages global interactions between semantic features at the end of the encoder to explicitly model full contextual information. Secondly, we introduce cross-attention in the skip connections to filter out non-semantic features, allowing a fine spatial recovery in the U-Net decoder.

Fig~\ref{fig:unet_rf}b) shows a cross-attention map induced by U-Transformer, which highlights the most important regions for segmenting the blue cross region in Fig~\ref{fig:unet_rf}a): our model leverages the long-range interactions with respect to other organs (liver, stomach, spleen) and their positions to properly segment the whole pancreas region, see Fig~\ref{fig:unet_rf}d).  
Quantitative experiments conducted on two abdominal CT-image datasets show the large performance gain brought out by U-Transformer compared to U-Net and to the local attention in~\cite{Oktay_MIDL_2018}.

\noindent\textbf{Related Work.} Transformers~\cite{Vaswani_NIPS_2017} have witnessed increasing success in the last five years, started in natural language processing with text embeddings~\cite{DBLP:journals/corr/abs-1810-04805}. A pioneer use of transformers in computer vision is non-local networks~\cite{Wang_CVPR_2018}, which combine self-attention with a convolutional backbone. Recent applications include object detection~\cite{Carion_ECCV_2020}, semantic segmentation~\cite{Ye_CVPR_2019,Wang_ECCV_2020}, and image classification~\cite{Dosovitskiy_ICLR_2021}. Recent works also focus on approximating self-attention mechanisms~\cite{Hu_RAL_2020,xiong2021nystromformer,Khan_arXiv_2021} to circumvent the high memory demand in transformers. All these approaches limit the use of transformers to self-attention. In contrast, U-Transformer use both self- and cross-attention, the latter being leveraged for improving the recovery of fine spatial and semantic information.

Attention models for medical image segmentation have also been used recently~\cite{DAF-18,ASDNet-18,SEE-18,Oktay_MIDL_2018,MS-Dual-20}. 
\cite{DAF-18,10.1007/978-3-030-12029-0_28} create attention  maps combining local and global features with a simple  attention module, and \cite{MS-Dual-20} successfully applies the Dual attention network in \cite{Fu_2019_CVPR} in different segmentation contexts. Despite the relevance of these works, they do not leverage the recent improvements obtained by transformers to model full range interactions. Attention U-Net~\cite{Oktay_MIDL_2018} uses gating signal in the skip connection, which acts as cross-attention. However, the attention weight maps are computed from local information only. In contrast, our cross-attention module incorporates rich region interactions and spatial information.

\vspace{-0.2cm}
\section{The U-Transformer Network}
\vspace{-0.1cm}

\begin{figure}[!ht]
    \centering
    \includegraphics[width=\textwidth]{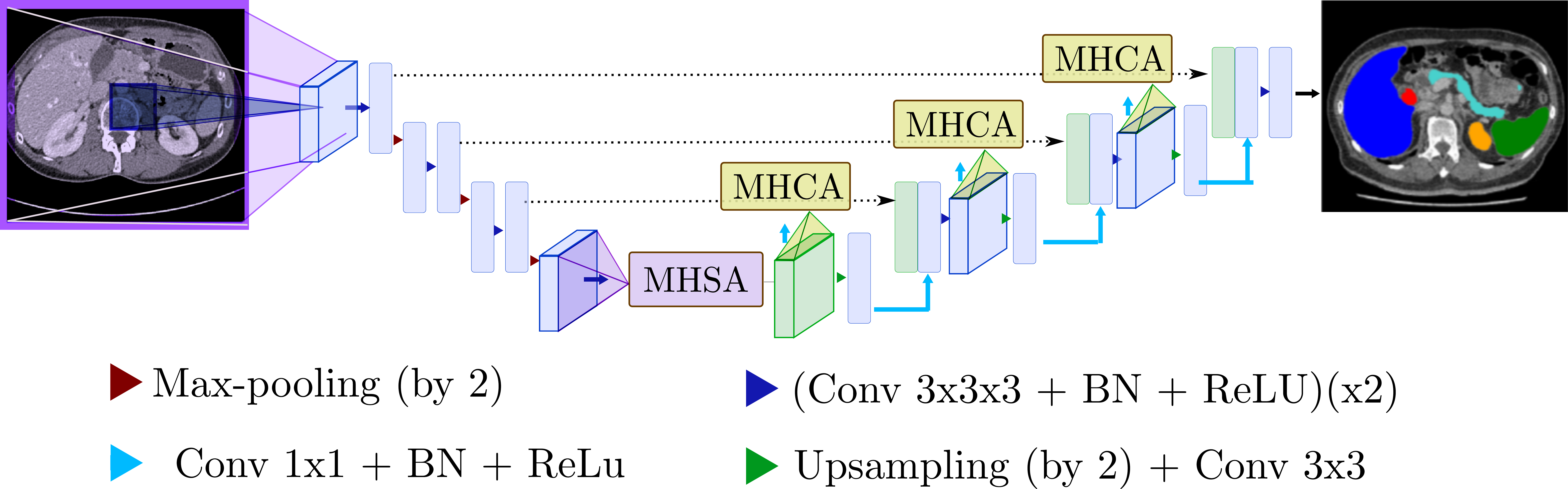}
        \caption{\textbf{U-Transformer} augments U-Nets with transformers to model long-range contextual interactions. The Multi-Head Self-Attention (MHSA) module at the end of the U-Net encoder gives access to a receptive field containing  the whole image (shown in purple), in contrast to the limited U-Net receptive field (shown in blue). Multi-Head Cross-Attention (MHCA) modules are dedicated to combine the semantic richness in high level feature maps with the high resolution ones coming from the skip connections.}
    \label{fig:unet_former_schema}
    \vspace{-0.5cm}
\end{figure}

As mentioned in Section~\ref{section:intro}, encoder-decoder U-shaped architectures lack global context information to handle complex medical image segmentation tasks.
We introduce the U-Transformer network, which augments U-Nets with attention modules built from multi-head transformers. U-Transformer models long-range contextual interactions and spatial dependencies by using two types of attention modules (see Fig~\ref{fig:unet_former_schema}): Multi-Head Self-Attention (MHSA) and Multi-Head Cross-Attention (MHCA). Both modules are designed to express a new representation of the input based on its self-attention in the first case (\textit{cf.} \ref{subsec:self-attention}) or on the attention paid to higher level features in the second (\textit{cf.} \ref{subsec:cross-attention}).

\subsection{Self-attention}
\label{subsec:self-attention}

The MHSA module is designed to extract long range structural information from the images. To this end, it is composed of multi-head self-attention functions as described in \cite{Vaswani_NIPS_2017} positioned at the bottom of the U-Net as shown in Figure~\ref{fig:unet_former_schema}. The main goal of MHSA is to connect every element in the highest feature map with each other, thus giving access to a receptive field including all the input image. The decision for one specific pixel can thus be influenced by any input pixel. The attention formulation is given in Equation~\ref{eq:self_attention}. A self-attention module takes three inputs, a matrix of queries $\V Q \in \mathbb{R}^{n\times d_k}$, a matrix of keys $\V K \in  \mathbb{R}^{n\times d_k}$ and a matrix of values $\V V \in  \mathbb{R}^{n\times d_k}$.

\vspace{-0.2cm}
\begin{equation}
    \label{eq:self_attention}
    \text{Attention}(\V Q,\V K,\V V) = \text{softmax}(\frac{\V Q \V K^T}{\sqrt{d_k}})\V V = \V A \V V
\end{equation}

\begin{figure}
    \centering
    \includegraphics[width=\textwidth]{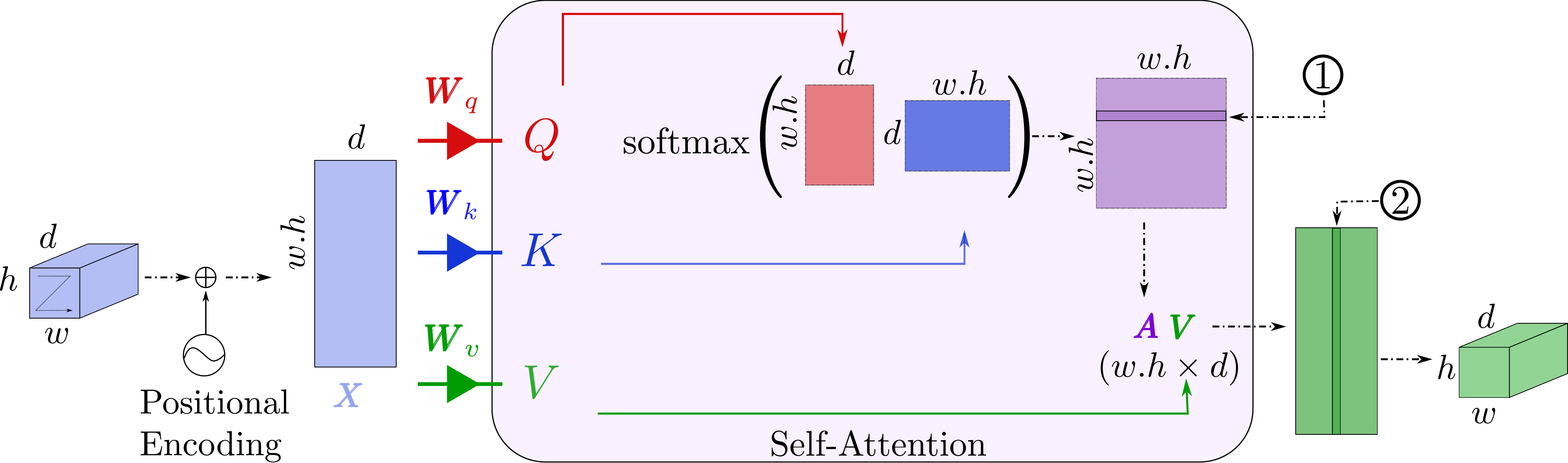}
    \caption{\textbf{MHSA module}: the input tensor is embedded into a matrix of queries $\V Q$, keys $\V K$ and values $\V V$. The attention matrix $\V A$ in purple is computed based on $\V Q$ and $\V K$. (1) A line of $\V A$ corresponds to the attention given to all the elements in $\V K$ with respect to one element in $\V Q$. (2) A column of the value $\V V$ corresponds to a feature map weighted by the attention in $\V A$.
    }
    \label{fig:self-attention}
    \vspace{-0.5cm}
\end{figure}

A line of the attention matrix $\V A \in \mathbb{R}^{n\times n}$ corresponds to the similarity of a given element in $\V Q$ with respect to all the elements in $\V K$. Then, the attention function performs a weighted average of the elements of the value $\V V$ to account for all the interactions between the queries and the keys as illustrated in Figure \ref{fig:self-attention}. In our segmentation task, $\V Q$, $\V K$ and $\V V$ share the same size and correspond to different learnt embedding of the highest level feature map denoted by $\V X$ in Figure \ref{fig:self-attention}. The embedding matrices are denoted as $\V W_q$, $\V W_k$ and $\V W_v$. The attention is calculated separately in multiple heads before being combined through another embedding. Moreover, to account for absolute contextual information, a positional encoding is added to the input features. It is especially relevant for medical image segmentation, where the different anatomical structures follow a fixed spatial position. The positional encoding can thus be leveraged to capture absolute and relative position between organs in MHSA.

\subsection{Cross-attention}
\label{subsec:cross-attention}

\begin{figure}
    \centering
    \includegraphics[width=\textwidth]{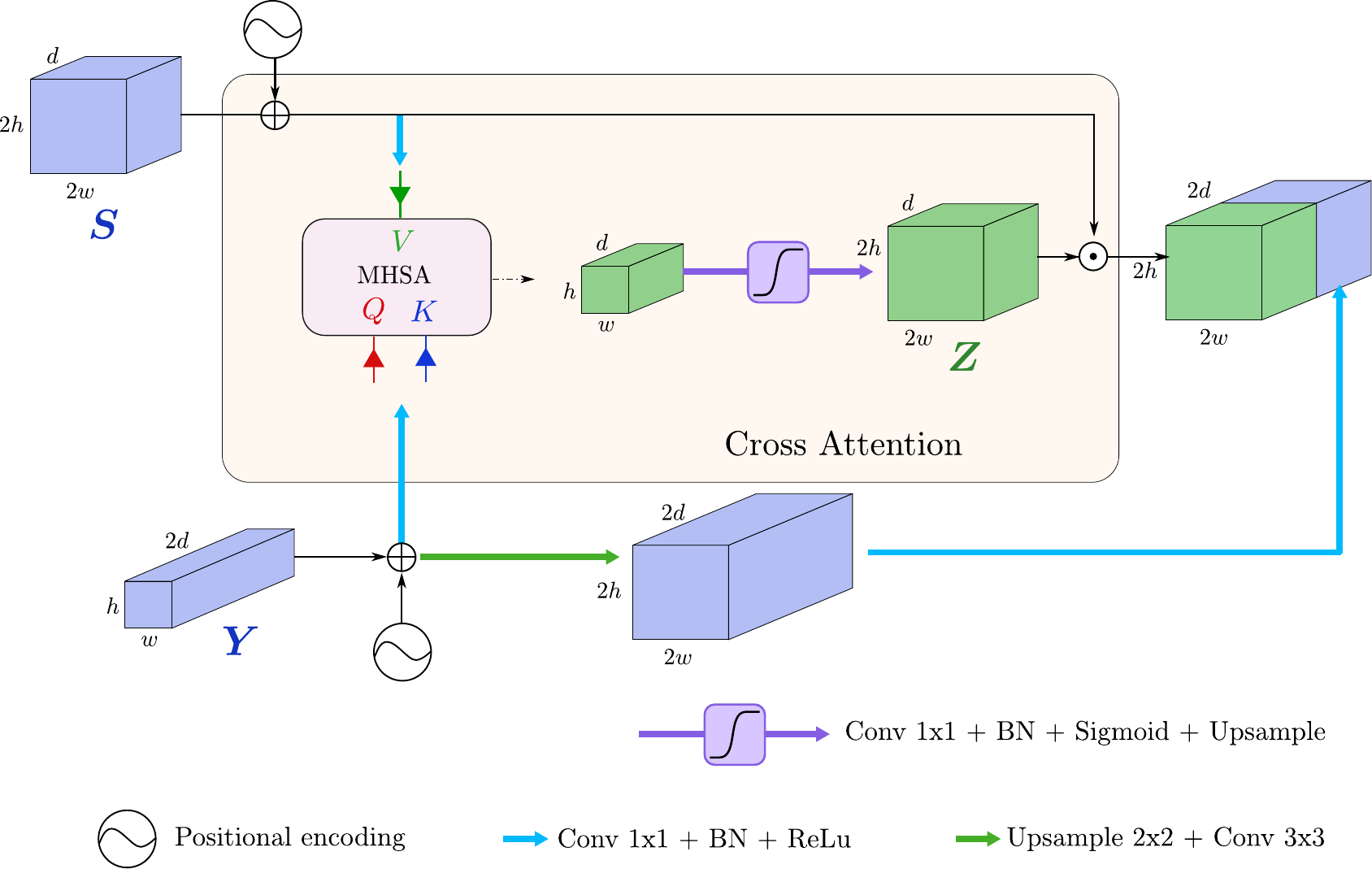}
    \caption{\textbf{MHCA module}: the value of the attention function corresponds to the skip connection $\V S$ weighted by the attention given to the high level feature map $\V Y$. This output is transformed into a filter $\V Z$ and applied to the skip connection.}
    \label{fig:MHCA}
    \vspace{-0.5cm}
\end{figure}

The MHSA module allows to connect every element in the input with each other. Attention may also be used to increase the U-Net decoder efficiency and in particular enhance the lower level feature maps that are passed through the skip connections. Indeed, if these skip connections insure to keep a high resolution information they lack the semantic richness that can be found deeper in the network. The idea behind the MHCA module is to turn off irrelevant or noisy areas from the skip connection features and highlight regions that present a significant interest for the application.
Figure~\ref{fig:MHCA} shows the cross-attention module. The MHCA block is designed as a gating operation of the skip connection $\V S$ based on the attention given to a high level feature map $\V Y$. The computed weight values are then re-scaled between 0 and 1 through a sigmoid activation function. The resulting tensor, denoted $\V Z$ in Figure~\ref{fig:MHCA}, is a filter where low magnitude elements indicate noisy or irrelevant areas to be reduced. A cleaned up version of $\V S$ is then given by the Hadamard product $ \V Z \odot \V S  $. Finally, the result of this filtering operation is concatenated with the high level feature tensor $\V Y$.

\section{Experiments}

We evaluate U-Transformer for abdominal organ segmentation on the TCIA pancreas public dataset, and an internal multi-organ dataset.

Accurate pancreas segmentation is particularly difficult, due to its small size, complex and variable shape, and because of the low contrast with the neighboring structures, see Fig~\ref{fig:unet_rf}. In addition, the multi-organ setting assesses how U-transformer can leverage attention from multi-organ annotations. 

\noindent\textbf{Experimental setup} The TCIA pancreas dataset\footnote{\url{https://wiki.cancerimagingarchive.net/display/Public/Pancreas-CT}} contains 82 CT-scans with pixel-level annotations. Each CT-scan has around $181 \sim 466$ slices of $512 \times 512$ pixels and a voxel spacing of ([$0.66 \sim 0.98$] $\times$ [$0.66 \sim 0.98$] $\times$ [$0.5 \sim 1.0$]) mm$^3$.

We also experiment with an Internal Multi-Organ (IMO) dataset composed of 85 CT-scans annotated with 7 classes: liver, gallbladder, pancreas, spleen, right and left kidneys, and stomach. Each CT-scan has around $57 \sim 500$ slices of $512 \times 512$ pixels and a voxel spacing of ([$0.42 \sim 0.98$] $\times$ [$0.42 \sim 0.98$] $\times$ [$0.63 \sim 4.00$])mm$^3$.

All experiments follow a 5-fold cross validation, using 80\% of images in training and 20\% in test. We use the Tensorflow library to train the model, with Adam optimizer ($10^{-4}$ learning rate, exponential decay scheduler).

We compare U-Transformer to the U-Net baseline~\cite{unet} and Attention U-Net~\cite{Oktay_MIDL_2018} with the same convolutional backbone for fair comparison. We also report performances with self-attention only (MHSA, section \ref{subsec:self-attention}), and the cross-attention only (MHCA, section \ref{subsec:cross-attention}). U-Net has $\sim30$M parameters, the overhead from U-transformer is limited (MHSA $\sim 5$M, each MHCA block $\sim 2.5$M).  

\subsection{U-Transformer performances}

Table~\ref{tab:overall_results} reports the performances in Dice averaged over the 5 folds, and over organs for IMO. 
U-Transformer outperforms U-Net by 2.4pts on TCIA and 1.3pts for IMO, and Attention U-Net by 1.7pts for TCIA and 1.6pts for IMO. The gains are consistent on all folds, and paired t-tests show that the improvement is significant with $p-$values $<3\%$ for every experiment.

\vspace{-0.5cm}
\begin{table}[]
    \centering
    \caption{Results for each method in Dice similarity coefficient (DSC, \%)} 
    \label{tab:overall_results}
    \resizebox{\columnwidth}{!}{%
    \begin{tabular}{l|cc|ccc}
        \toprule[1pt]
        Dataset & U-Net~\cite{unet} & Attn U-Net~\cite{Oktay_MIDL_2018} & MHSA & MHCA & U-Transformer \\
        \hline
        TCIA & 76.13 ($\pm$ 0.94) & 76,82 ($\pm$ 1.26) & 77.71 ($\pm$ 1.31) & 77.84 ($\pm$ 2.59) & \textbf{78.50} ($\pm$ 1.92) \\
        IMO & 86.78 ($\pm$ 1.72) & 86.45 ($\pm$ 1.69) & 87.29 ($\pm$ 1.34) & 87.38 ($\pm$ 1.53) & \textbf{88.08} ($\pm$ 1.37) \\
        \toprule[1pt]
    \end{tabular}
    }
    \vspace{-0.3cm}
\end{table}

\begin{figure}[!ht]
    \centering
    \includegraphics[width=\textwidth]{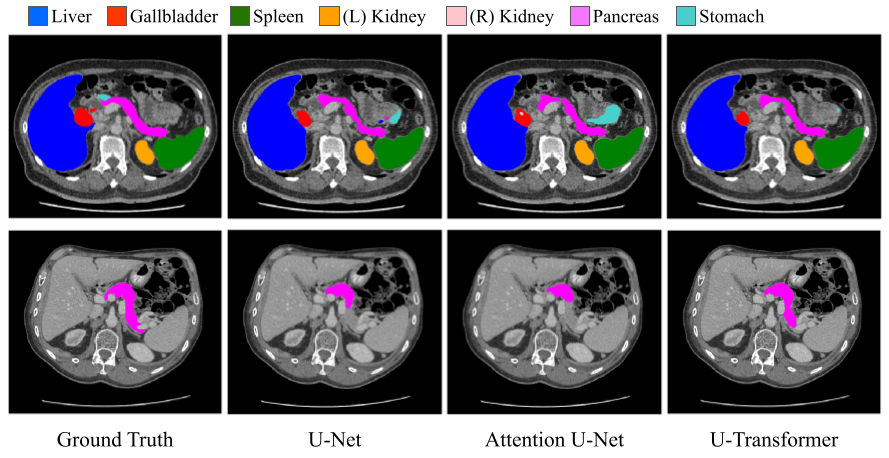}
    \caption{Segmentation results for U-Net~\cite{unet}, Attention U-Net~\cite{Oktay_MIDL_2018} and U-Transformer on the multi-organ IMO dataset (first row) and on TCIA pancreas (second row).}
    \label{fig:segmentation_results}
    \vspace{-0.5cm}
\end{figure}

Figure~\ref{fig:segmentation_results} provides qualitative segmentation comparison between U-Net, Attention U-Net and U-Transformer. We observe that U-Transformer performs better on difficult cases, where the local structures are ambiguous. For example, in the second row, the pancreas has a complex shape which is missed by U-Net and Attention U-Net but U-Transformer successfully segments the organ.

In Table~\ref{tab:overall_results}, we can see that the self-attention (MHSA) and cross-attention (MCHA) alone already outperform U-Net and Attention U-Net on TCIA and IMO. Since MCHA and Attention U-Net apply attention mechanisms at the skip connection level, it highlights the superiority of modeling global interactions between anatomical structures 
and positional information instead of the simple local attention in~\cite{Oktay_MIDL_2018}. Finally, the combination of MHSA and MHCA in U-Transformer shows that the two attention mechanisms are complementary and can collaborate to provide better segmentation predictions.

Table~\ref{tab:multiorgan_per_patient} details the results for each organ on the multi-organ IMO dataset. This further highlights the interest of U-Transformer, which significantly outperforms U-Net and Attention U-Net for the most challenging organs: pancreas: +3.4pts, gallbladder: +1.3pts and stomach: +2.2pts. This validates the capacity of U-Transformer to leverage multi-label annotations to drive the interactions between anatomical structures, and use easy organ predictions to improve the detection and delineation of more difficult ones. We can note that U-Transformer is better for every organ, even the liver which has a high score $>95\%$ with U-Net.

\vspace{-0.5cm}
\begin{table}[h!]
    \centering
    \caption{Results on IMO in Dice similarity coefficient (DSC, \%) detailed per organ.}
    \label{tab:multiorgan_per_patient}
    \resizebox{\columnwidth}{!}{%
    \begin{tabular}{l|cc|ccc}
        \toprule[1pt]
        Organ & U-Net~\cite{unet} & Attn U-Net~\cite{Oktay_MIDL_2018} & MHSA & MHCA & U-Transformer \\
        \hline
        Pancreas & 69.71 \footnotesize{($\pm$ 3.74)} & 68.65 \footnotesize{($\pm$ 2.95)} & 71.64 \footnotesize{($\pm$ 3.01)} & 71.87 \footnotesize{($\pm$ 2.97)} & \textbf{73.10} \footnotesize{($\pm$ 2.91)} \\
        Gallbladder & 76.98 \footnotesize{($\pm$ 6.60)} & 76.14 \footnotesize{($\pm$ 6.98)} & 76.48 \footnotesize{($\pm$ 6.12)} & 77.36 \footnotesize{($\pm$ 6.22)} & \textbf{78.32} \footnotesize{($\pm$ 6.12)} \\
        Stomach & 83.51 \footnotesize{($\pm$ 4.49)} & 82.73 \footnotesize{($\pm$ 4.62)} & 84.83 \footnotesize{($\pm$ 3.79)} & 84.42 \footnotesize{($\pm$ 4.35)} & \textbf{85.73} \footnotesize{($\pm$ 3.99)} \\
        Kidney(R) & 92.36 \footnotesize{($\pm$ 0.45)} & 92.88 \footnotesize{($\pm$ 1.79)} & 92.91 \footnotesize{($\pm$ 1.84)} & 92.98 \footnotesize{($\pm$ 1.70)} & \textbf{93.32} \footnotesize{($\pm$ 1.74)} \\
        Kidney(L) & 93.06 \footnotesize{($\pm$ 1.68)} & 92.89 \footnotesize{($\pm$ 0.64)} & 92.95 \footnotesize{($\pm$ 1.30)} & 92.82 \footnotesize{($\pm$ 1.06)} & \textbf{93.31} \footnotesize{($\pm$ 1.08)} \\
        Spleen & 95.43 \footnotesize{($\pm$ 1.76)} & 95.46 \footnotesize{($\pm$ 1.95)} & 95.43 \footnotesize{($\pm$ 2.16)} & 95.41 \footnotesize{($\pm$ 2.21)} & \textbf{95.74} \footnotesize{($\pm$ 2.07)} \\
        Liver & 96.40 \footnotesize{($\pm$ 0.72)} & 96.41 \footnotesize{($\pm$ 0.52)} & 96.82 \footnotesize{($\pm$ 0.34)} & 96.79 \footnotesize{($\pm$ 0.29)} & \textbf{97.03} \footnotesize{($\pm$ 0.31)} \\
        \toprule[1pt]
    \end{tabular}
    }
    \vspace{-0.3cm}
\end{table}

\subsection{U-Transformer analysis and properties}

\noindent\textbf{Positional encoding and multi-level MHCA.} The Positional Encoding (PE) allows to leverage the absolute position of the objects in the image. Table~\ref{tab:potional_encoding} shows an analysis of its impact, on one fold on both datasets. For MHSA, the PE improves the results by +0.7pt for TCIA and +0.6pt for IMO. For MHCA, we evaluate a single level of attention with and without PE. We can observe an improvement of +1.7pts for TCIA and +0.6pt for IMO between the two versions.

\begin{figure}[t]
    \centering
    \includegraphics[width=\textwidth]{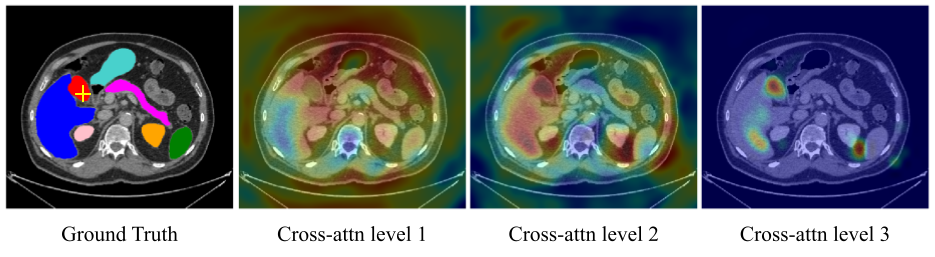}
    \caption{Cross-attention maps for the yellow-crossed pixel (left image).}
    \label{fig:cross_attention}
    \vspace{-0.5cm}
\end{figure}

Table~\ref{tab:potional_encoding} also shows the favorable impact of using multi \textit{vs} single-level attention for MHCA: +1.8pts for TCIA and +0.6pt for IMO. It is worth noting that Attention U-Net uses multi-level attention but remains below MHCA with a single level. Figure~\ref{fig:cross_attention} shows attention maps at each level of U-Transformer: level 3 corresponds to high-resolution features maps, and tends to focus on more specific regions compared to the first levels.

\vspace{-0.5cm}
\begin{table}[]
    \centering
    \caption{Ablation study on the positional encoding and multi-level on one fold of TCIA and IMO.}
    \label{tab:potional_encoding}
    \resizebox{\columnwidth}{!}{%
    \begin{tabular}{lcc|cc|ccc}
        \toprule[1pt]
        & & & \multicolumn{2}{c|}{MHSA} & \multicolumn{3}{c}{MHCA} \\
         & U-Net & Attn U-Net & wo PE -- & w PE & 1 lvl wo PE -- & 1 lvl w PE -- & multi-lvl w PE \\
        TCIA & 76.35 & 77.23 & 78.17 & \textbf{78.90} & 77.18 & 78.88 & \textbf{80.65} \\
        IMO & 88.18 & 87.52 & 88.16 & \textbf{88.76} & 87.96 & 88.52 & \textbf{89.13} \\
        \toprule[1pt]
    \end{tabular}
    }
    \vspace{-0.3cm}
\end{table}

\noindent\textbf{Further analysis.} To further analyse the behaviour of U-Transformer, we evaluate the impact of the number of attention heads for MHSA (appendix, Figure~\ref{fig:number_of_heads}): more heads lead to better performances, but the biggest gain comes from the first head (\ie U-Net to MHSA). Finally, the evaluation of U-Transformer with respect to the Hausdorff distance (appendix, Table~\ref{tab:hd_scores}) follows the same trend than with Dice score. This highlights the capacity of U-Transformer to reduce prediction artefacts by means of self- and cross-attention.

\section{Conclusion}

This paper introduces the U-Transformer network, which augments a U-shaped FCN with Transformers. We propose to use self and cross-attention modules to model long-range interactions and spatial dependencies. We highlight the relevance of the approach for abdominal organ segmentation, especially for small and complex organs. Future works could include the study of U-Transformer in 3D networks, with other modalities such as MRI or US images, as well as for other medical image tasks.

%
%
%
\bibliographystyle{splncs04}
\bibliography{bibliography.bib}

\appendix

\begin{figure}[!h]
    \centering
    \includegraphics[width=\textwidth]{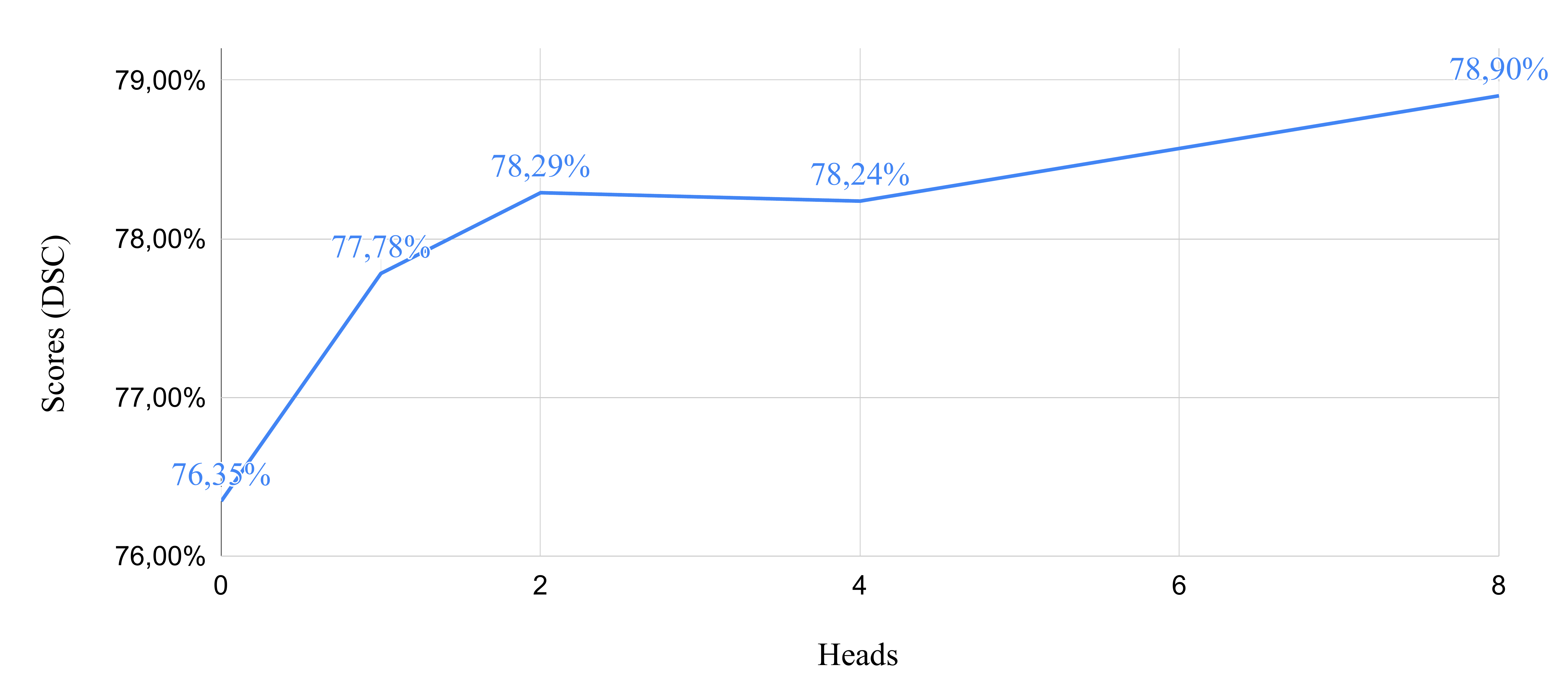}
    \caption{Evolution of the Dice Score on TCIA (fold 1) when the number of heads varies between 0 and 8 in MHSA. }
    \label{fig:number_of_heads}
\end{figure}

\begin{table}[h!]
    \centering
    \caption{Hausdorff Distances (HD) for the different models} 
    \label{tab:hd_scores}
    \begin{tabular}{lccc}
        \toprule[1pt]
        Dataset & U-Net & Attn U-Net & U-Transformer \\
        \hline
        TCIA & 13.61 ($\pm$ 2.01) & 12.48 ($\pm$ 1.36) & \textbf{12.34} ($\pm$ 1.51) \\
        IMO & 12.06 ($\pm$ 1.65) & 12.13 ($\pm$ 1.58) & \textbf{12.00} ($\pm$ 1.32) \\
        \toprule[1pt]
    \end{tabular}
\end{table}

\end{document}